\documentclass[italian,english]{article}
\usepackage[T1]{fontenc}
\usepackage[latin1]{inputenc}
\usepackage{color}
\usepackage{graphicx}
\usepackage{amssymb}

\makeatletter

 \newcommand{\lyxaddress}[1]{
   \par {\raggedright #1 
   \vspace{1.4em}
   \noindent\par}
 }

\usepackage{babel}
\makeatother
\begin{document}

\title{\textbf{The Virgo - MiniGRAIL cross correlation for the detection
of scalar gravitational waves}}

\author{\textbf{Christian Corda}}

\maketitle

\lyxaddress{\begin{center}INFN - Sezione di Pisa and Università di Pisa, Via
F. Buonarroti 2, I - 56127 PISA, Italy\end{center}}

\lyxaddress{\begin{center}\textit{E-mail address:} \textcolor{blue}{christian.corda@ego-gw.it} \end{center}}

\begin{abstract}
After a review of the frequency - dependent angular pattern of interferometers
in the TT gauge for scalar gravitational waves (SGWs), which has been
recently analysed by Capozziello and Corda, in this letter the result
is used to study the cross correlation between the Virgo interferometer
and the MiniGRAIL resonant sphere. It is shown that the overlap reduction
function for the cross correlation bewteen Virgo and the monopole
mode of MiniGRAIL is very small, but a maximum is also found in the
correlation at about $2710Hz$, in the range of the MiniGRAIL sensitivity. 
\end{abstract}

\lyxaddress{PACS numbers: 04.80.Nn, 04.30.Nk, 04.50.+h}

The design and construction of a number of sensitive detectors for
gravitational waves (GWs) is underway today. There are some laser
interferometers and many bar detectors currently in operation  and
several interferometers and bars are in a phase of planning and proposal
stages (for the current status of gravitational waves experiments
see \cite{key-1,key-2,key-3,key-4,key-5}).

The results of these detectors will have a fundamental impact on astrophysics
and gravitation physics. There will be lots of experimental data to
be analyzed, and theorists will be forced to interact with lots of
experiments and data analysts to extract the physics from the data
stream.

Detectors for GWs will also be important to verify that GWs only change
distances perpendicular to their direction of propagation and to confirm
or ruling out the physical consistency of General Relativity or of
any other theory of gravitation \cite{key-6,key-7,key-8,key-9,key-10}.

These detectors are in principle sensitive also to a hypotetical \textit{scalar}
component of gravitational radiation, that appears in extended theories
of gravity like scalar-tensor gravity \cite{key-11,key-12,key-13,key-14}
and high order theories \cite{key-7,key-8,key-9,key-10}.

While the response of interferometers to tensorial waves has been
calculated in lots of works (see for example \cite{key-15,key-16,key-17}),
the coupling between interferometers and scalar waves is a more recent
field of interest and it has not an analogous number of works in the
literature. Only recently, Capozziello and Corda have shown that the
response of an interferometer to massless SGWs is invariant in three
different gauges known in literature in its full frequency dependence
\cite{key-6} while previous works started from the assumption that
the wavelength of the SGW is much larger than the distance between
the test masses (see \cite{key-12,key-13,key-14} for examples).

In this letter, after a review of the frequency - dependent angular
pattern of interferometers in the TT gauge, which is due to provide
a context to bring out the relevance of the argument, the result will
be used to study a cross correlation between the Virgo interferometer
and the MiniGRAIL resonant sphere \cite{key-18,key-19} generalizing
the analysis of \cite{key-13} for a real detector.

Following \cite{key-6}, in the TT gauge, for a purely massless SGW
propagating in the positive $z$ direction, with the interferometer
located at the origin of the coordinate sistems with arms in the $\overrightarrow{x}$
and $\overrightarrow{y}$ directions, the metric perturbation is given
by (see \cite{key-6,key-12,key-13}, and note that we work with $c=1$
and $\hbar=1$ in this letter):

\begin{equation}
h_{\mu\nu}(t-z)=\Phi(t-z)e_{\mu\nu}^{(s)},\label{eq: perturbazione scalare}\end{equation}

where $\Phi\ll1$, $e_{\mu\nu}^{(s)}\equiv diag(0,1,1,0)$, and the
line element is

\begin{equation}
ds^{2}=-dt^{2}+dz^{2}+[1+\Phi(t-z)][dx^{2}+dy^{2}].\label{eq: metrica puramente scalare}\end{equation}

To compute the response function for an arbitrary propagating direction
of the SGW one recalls that the arms of the interferometer are in
the $\overrightarrow{u}$ and $\overrightarrow{v}$ directions, while
the $x,y,z$ frame of (\ref{eq: metrica puramente scalare}) is adapted
to the propagating SGW. Thus a spatial rotation of the coordinate
system is needed:

\begin{equation}
\begin{array}{ccc}
u & = & -x\cos\theta\cos\phi+y\sin\phi+z\sin\theta\cos\phi\\
\\v & = & -x\cos\theta\sin\phi-y\cos\phi+z\sin\theta\sin\phi\\
\\w & = & x\sin\theta+z\cos\theta,\end{array}\label{eq: rotazione}\end{equation}

or, in terms of the $x,y,z$ frame:

\begin{equation}
\begin{array}{ccc}
x & = & -u\cos\theta\cos\phi-v\cos\theta\sin\phi+w\sin\theta\\
\\y & = & u\sin\phi-v\cos\phi\\
\\z & = & u\sin\theta\cos\phi+v\sin\theta\sin\phi+w\cos\theta.\end{array}\label{eq: rotazione 2}\end{equation}

In this way the SGW is propagating from an arbitrary direction to
the interferometer (see figure 1). %
\begin{figure}
\includegraphics{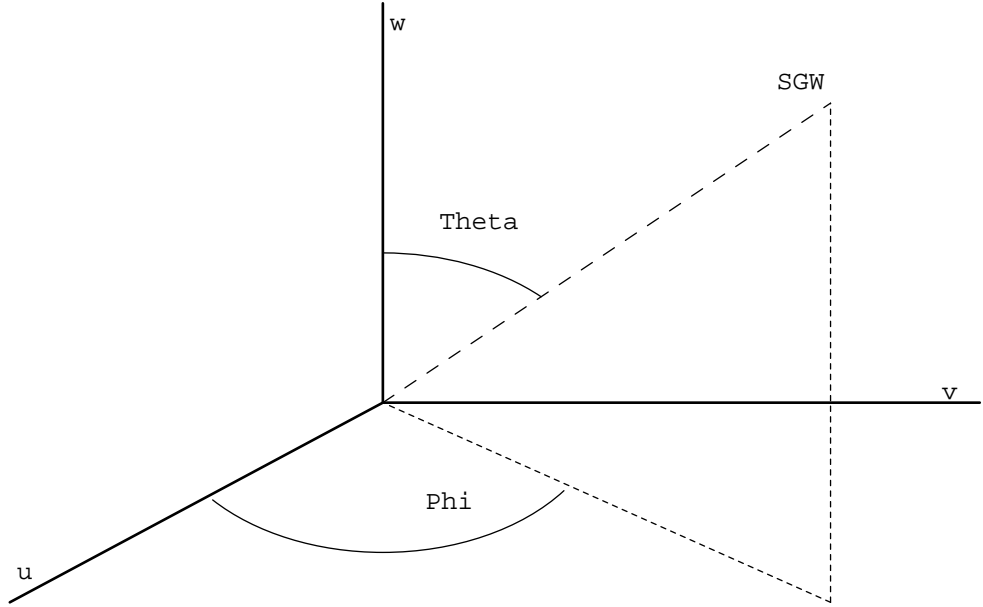}

\caption{a SGW propagating from an arbitrary direction}
\end{figure}

The beam splitter is also put in the origin of the new coordinate
system (i.e. $u_{b}=0$, $v_{b}=0$). 

The transformation for the metric tensor is \cite{key-20}:

\begin{equation}
g^{ik}=\frac{\partial x^{i}}{\partial x'^{l}}\frac{\partial x^{k}}{\partial x'^{m}}g'^{lm}.\label{eq: trasformazione metrica}\end{equation}

By using eq. (\ref{eq: rotazione}), eq. (\ref{eq: rotazione 2})
and eq. (\ref{eq: trasformazione metrica}), the line element (\ref{eq: metrica puramente scalare})
in the $\overrightarrow{u}$ direction becomes:

\begin{equation}
ds^{2}=-dt^{2}+[1+(1-\sin^{2}\theta\cos^{2}\phi)\Phi(t-u\sin\theta\cos\phi)]du^{2}.\label{eq: metrica scalare lungo u 2}\end{equation}

A good way to analyze variations in the proper distance (time) is
by means of {}``bouncing photons'' (see \cite{key-6,key-17} and
figure 2). 

\begin{figure}
\includegraphics{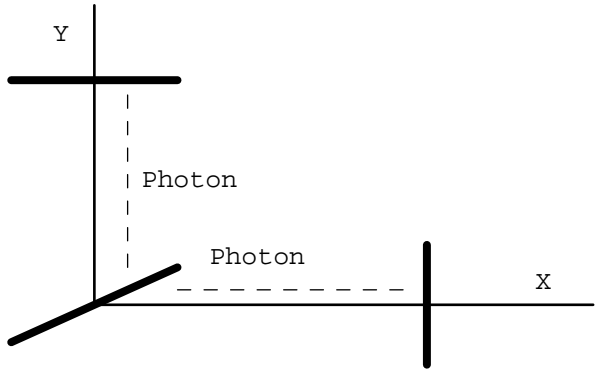}

\caption{photons can be launched from the beam-splitter to be bounced back
by the mirror}
\end{figure}

Let us compute the variaton of the proper distance that a photon covers
to make  a round-trip from the beam-splitter to the mirror of an interferometer
with the line-element choice (\ref{eq: metrica scalare lungo u 2}). 

A special property of the TT gauge is that an inertial test mass initially
at rest in these coordinates, remains at rest throughout the entire
passage of the SGW \cite{key-6,key-12,key-13}. Here we have to clarify
the use of words {}`` at rest'': we want to mean that the coordinates
of the test masses do not change in the presence of the SGW. The proper
distance between the beam-splitter and the mirror of the interferometer
changes even though their coordinates remain the same \cite{key-6,key-13}. 

Putting the condition for null geodesics $ds^{2}=0$ in eq. (\ref{eq: metrica scalare lungo u 2})
the coordinate velocity of the photon, which is a convenient quantity
for calculations of the photon propagation time between the beam-splitter
and the mirror \cite{key-6,key-17}, is:

\begin{equation}
v^{2}\equiv(\frac{du}{dt})^{2}=\frac{1}{[1+(1-\sin^{2}\theta\cos^{2}\phi)\Phi(t-u\sin\theta\cos\phi)]}.\label{eq: velocita' fotone u}\end{equation}

Because the coordinates of the beam-splitter $u_{b}=0$ and of the
mirror $u_{m}=L$ do not change under the influence of the SGW, the
duration of the forward trip can be found as

\begin{equation}
T_{1}(t)=\int_{0}^{L}\frac{du}{v(t'-u\sin\theta\cos\phi)},\label{eq: durata volo}\end{equation}

with 

\begin{center}$t'=t-(L-u)$.\end{center}

In the last equation $t'$ is the retardation time (i.e. $t$ is the
time at which the photon arrives in the position $L$, so $L-u=t-t'$).

To first order in $\Phi$ this integral is approximated by

\begin{equation}
T_{1}(t)=T+\frac{1-\sin^{2}\theta\cos^{2}\phi}{2}\int_{0}^{L}\Phi(t'-u\sin\theta\cos\phi)du,\label{eq: durata volo andata approssimata u}\end{equation}

where

\begin{center}$T=L$ \end{center}

is the transit time of the photon in the absence of the SGW. Similiary,
the duration of the return trip will be\begin{equation}
T_{2}(t)=T+\frac{1-\sin^{2}\theta\cos^{2}\phi}{2}\int_{L}\Phi(t'-u\sin\theta\cos\phi)(-du),\label{eq: durata volo ritorno approssimata u}\end{equation}

though now the retardation time is 

\begin{center}$t'=t-u$.\end{center}

The round-trip time will be the sum of $T_{2}(t)$ and $T_{1}[t-T_{2}(t)]$.
The latter can be approximated by $T_{1}(t-T)$ because the difference
between the exact and the approximate values is second order in $\Phi$.
Then, to first order in $\Phi$, the duration of the round-trip will
be

\begin{equation}
T_{r.t.}(t)=T_{1}(t-T)+T_{2}(t).\label{eq: durata round trip}\end{equation}

By using eqs. (\ref{eq: durata volo andata approssimata u}) and (\ref{eq: durata volo ritorno approssimata u})
we see immediatly that deviations of this round-trip time (i.e. proper
distance) from its imperurbated value are given by

\begin{equation}
\begin{array}{c}
\delta T(t)=\frac{1-\sin^{2}\theta\cos^{2}\phi}{2}\int_{0}^{L}[\Phi(t-2T+u(1-\sin\theta\cos\phi))+\\
\\+\Phi(t-u(1+\sin\theta\cos\phi))]du.\end{array}\label{eq: variazione temporale in u}\end{equation}
Now, using the Fourier transform of the scalar field, defined by

\begin{equation}
\tilde{\Phi}(\omega)=\int_{-\infty}^{\infty}dt\Phi(t)\exp(i\omega t),\label{eq: trasformata di fourier}\end{equation}

the analysis can be transled in the frequency domain obtaining

\begin{equation}
\frac{\delta\tilde{T}(\omega)}{T}=\Upsilon_{u}(\omega)\tilde{\Phi}(\omega)\label{eq: segnale u}\end{equation}

where the total response function in the $\overrightarrow{u}$ direction
is given by:

\begin{equation}
\begin{array}{c}
\Upsilon_{u}(\omega)=\frac{1}{2i\omega L}[-1+\exp(2i\omega L)+\\
\\+\sin\theta\cos\phi((1+\exp(2i\omega L)-2\exp i\omega L(1+\sin\theta\cos\phi))].\end{array}\label{eq: risposta totale lungo u 2}\end{equation}

In the same way the line element (\ref{eq: metrica puramente scalare})
in the $\overrightarrow{v}$ direction becomes:

\begin{equation}
ds^{2}=-dt^{2}+[1+(1-\sin^{2}\theta\sin^{2}\phi)\Phi(t-v\sin\theta\sin\phi)]dv^{2},\label{eq: metrica + lungo v}\end{equation}

thus, with the same analysis of the $\overrightarrow{u}$ direction,
the total response function in the $\overrightarrow{v}$ direction
is obtained:

\begin{equation}
\begin{array}{c}
\Upsilon_{v}(\omega)=\frac{1}{2i\omega L}[-1+\exp(2i\omega L)+\\
\\+\sin\theta\sin\phi((1+\exp(2i\omega L)-2\exp i\omega L(1+\sin\theta\sin\phi))].\end{array}\label{eq: risposta totale lungo u 2}\end{equation}

Then, the total response function of an interferometer, defined by
$\tilde{H}(\omega)\equiv\Upsilon_{u}(\omega)-\Upsilon_{v}(\omega),$
is:

\begin{equation}
\begin{array}{c}
\tilde{H}(\omega)=\frac{\sin\theta}{2i\omega L}\{\cos\phi[1+\exp(2i\omega L)-2\exp i\omega L(1+\sin\theta\cos\phi)]+\\
\\-\sin\phi[1+\exp(2i\omega L)-2\exp i\omega L(1+\sin\theta\sin\phi)]\}.\end{array}\label{eq: risposta totale Virgo}\end{equation}
Eq. (\ref{eq: risposta totale Virgo}) is also in perfect agreement
with the detector pattern of \cite{key-13} and \cite{key-14} in
the low frequencies limit ($\omega\rightarrow0$):

\begin{equation}
\tilde{H}(\omega\rightarrow0)=-\sin^{2}\theta\cos2\phi.\label{eq: risposta totale approssimata}\end{equation}

In figs. 3 and 4 the absolute value of the total response function
of the Virgo interferometer ($L=3$ Km) ) for SGWs with $\theta=\frac{\pi}{4}$
and $\phi=\frac{\pi}{3}$, and the angular dependence of the response
of the Virgo interferometer for a SGW with a frequency of $f=100Hz$
are shown respectively.

\begin{figure}
\includegraphics{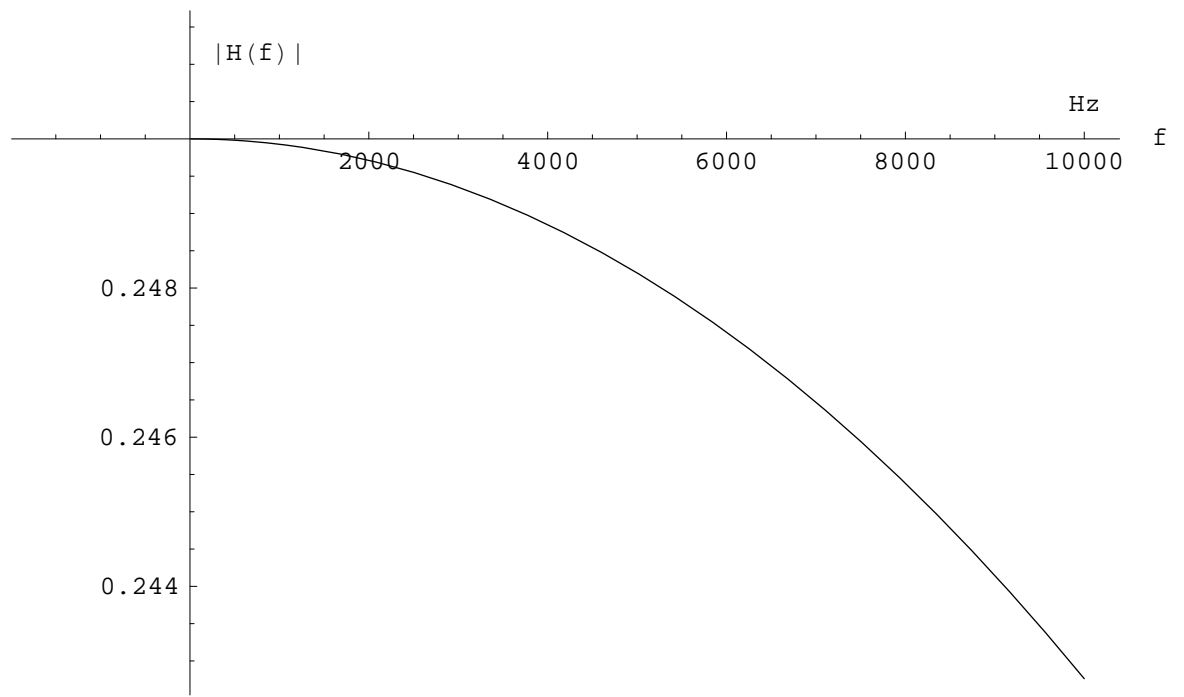}

\caption{the absolute value of the total response function of the Virgo interferometer
to SGWs for $\theta=\frac{\pi}{4}$ and $\phi=\frac{\pi}{3}$. }
\end{figure}

\begin{figure}
\includegraphics{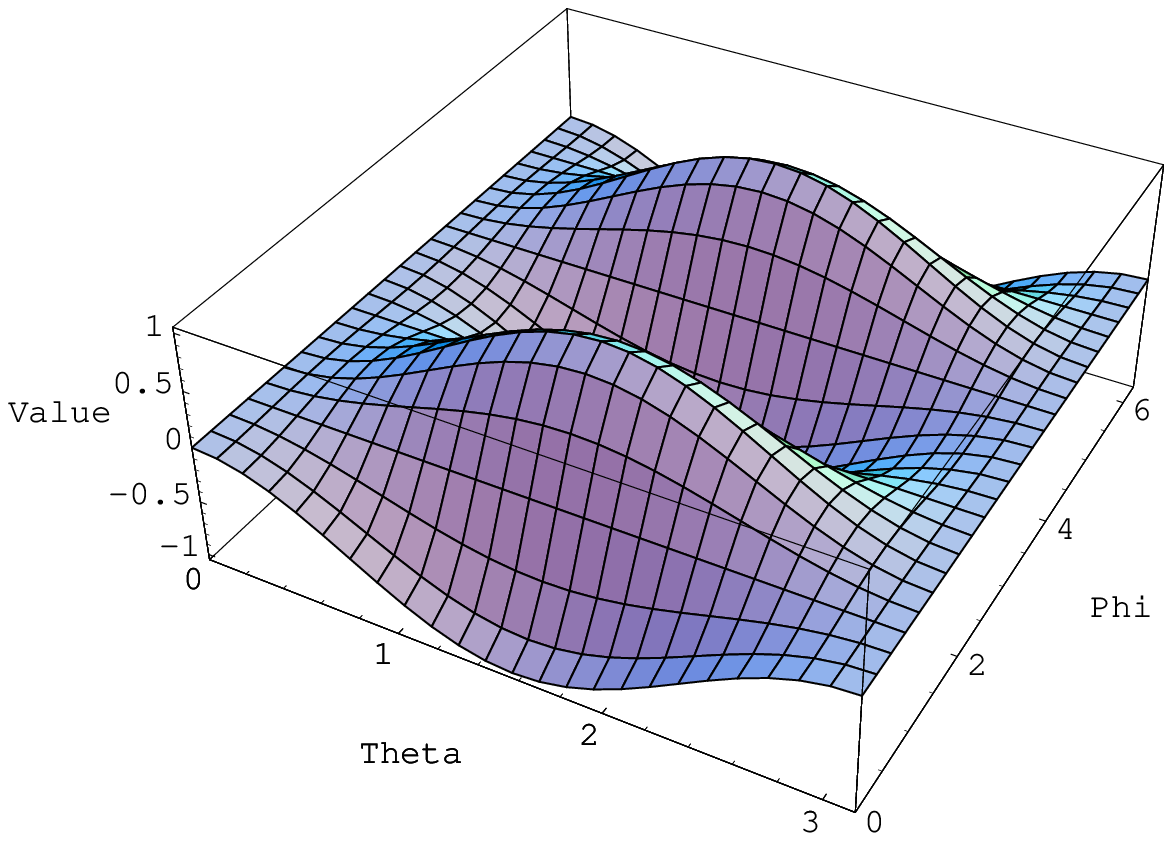}

\caption{the angular dependence of the response of the Virgo interferometer
for a SGW with a frequency of $f=100Hz$}
\end{figure}

Now, the analysis of \cite{key-13} , where a cross correlation between
the Virgo interferometer and a hypothetical resonant sphere in Frascati,
near Rome, was studied, will be generalized for a real detector, the
MiniGRAIL sphere \cite{key-18,key-19}.

In \cite{key-13} the overlap reduction function for SGWs has been
computed generalizing the well known Flanagan's overlap reduction
function for ordinary (tensorial) GWs defined in \cite{key-21}. The
overlap reduction function for SGWs has also been computed in \cite{key-22}.

In the cross correlation between a sphere and an interferometer, the
monopole mode of a resonant sphere is especially interesting, because
it cannot be excited by tensorial waves \cite{key-13,key-22}.

The authors of \cite{key-13} found the relation (see eq. (56) of
\cite{key-13} )

\begin{equation}
\Gamma(f)=(\sin^{2}\theta\cos2\phi)j_{2}(2\pi fd),\label{eq: overlap MN}\end{equation}

where $j_{2}$ is the second spherical Bessel function and $\theta$
and $\phi$ are the angular coordinates of the MINIGRAIL sphere with
respect the Virgo interferometer. (see figure 5). %
\begin{figure}
\includegraphics{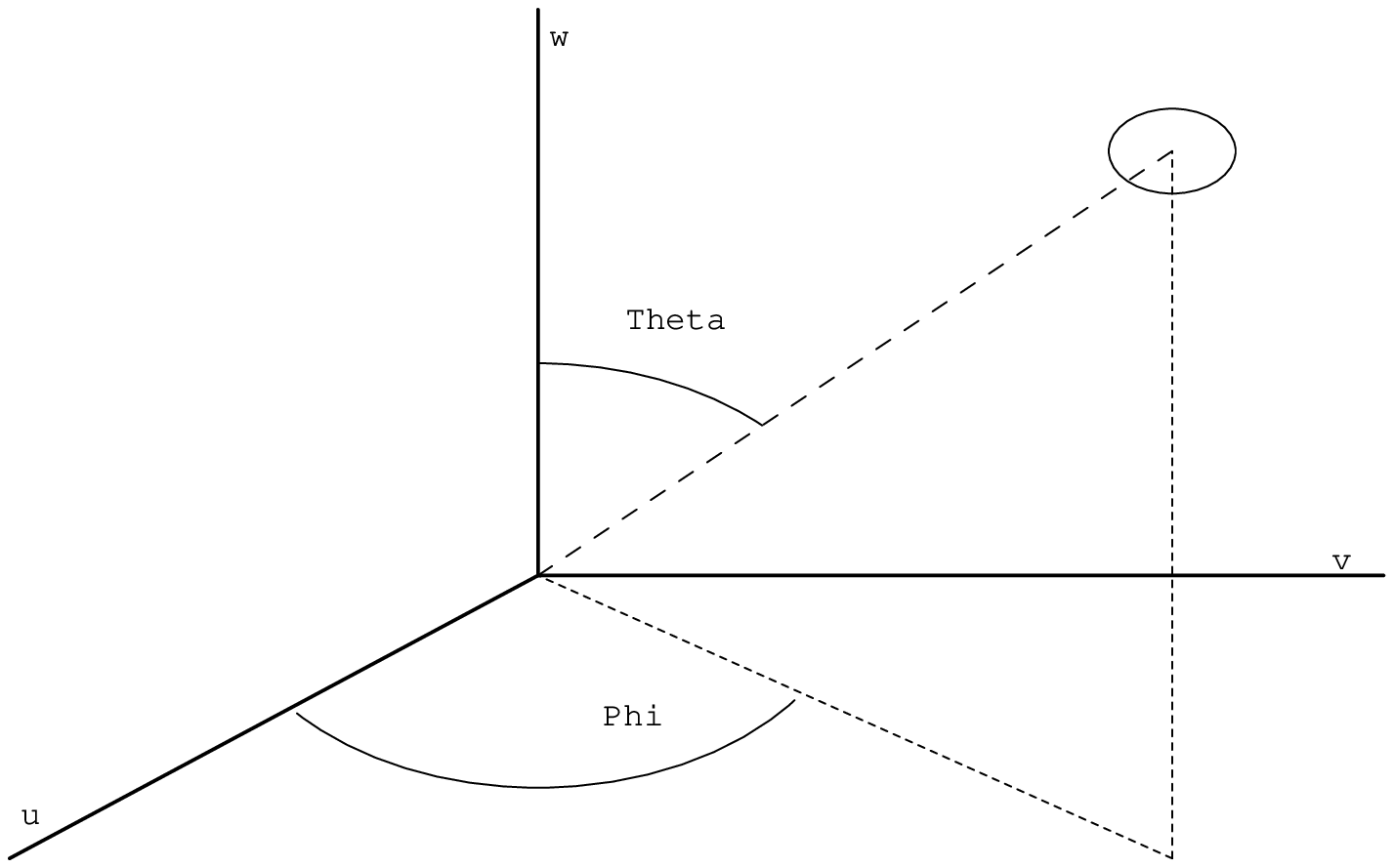}

\caption{The Virgo-MiniGRAIL cross correlation}
\end{figure}

In eq. (\ref{eq: overlap MN}) the low frequencies approximation (\ref{eq: risposta totale approssimata})
has been used. Replacing the low frequencies approximation with our
frequency dependent angular pattern (\ref{eq: risposta totale Virgo})
the more general overlap reduction function is obtained:

\begin{equation}
\begin{array}{c}
\Gamma(f)=\frac{-\sin\theta}{4i\pi fL}\{\cos\phi[1+\exp(4i\pi fL)-2\exp i2\pi fL(\sin\theta\cos\phi-1)]+\\
\\-\sin\phi[1+\exp(4i\pi fL)-2\exp i2\pi fL(\sin\theta\sin\phi-1)]\} j_{2}(2\pi fd).\end{array}\label{eq: overlap MN 2}\end{equation}
MiniGRAIL works in the frequency range between $2670-3130Hz$ \cite{key-18,key-19}
with a resonance frequency of about $2900Hz$, and, because the angular
pattern in the Virgo- MiniGRAIL direction is almost constant in this
range (figure 6), the function $j_{2}(2\pi fd)$ is a good approximation
for the frequency dependence of the overlap reduction function.

\begin{figure}
\includegraphics{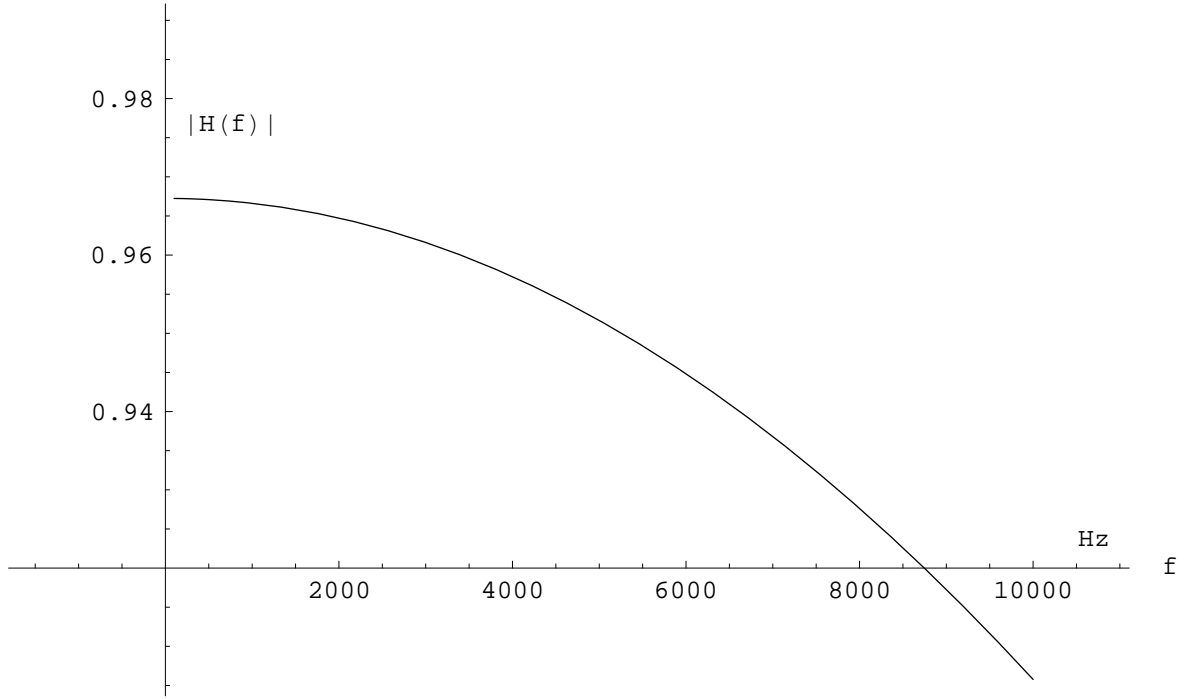}

\caption{the absolute value of the total response function of the Virgo interferometer
to SGWs for the Virgo- MiniGRAIL direction. }
\end{figure}

In figure 7 the frequency dependence of the overlap reduction function
for the approximate distance between the location of the Virgo interferometer
and the location of the MiniGRAIL resonant sphere in the MiniGRAIL
frequency range is shown. The location of MiniGRAIL is Lat. N 52.16,
Lon E 4.48, while the location of Virgo is Lat. N 43.63, Lon E 10.50
and the Virgo-MiniGRAIL distance is about 1090 kms. Figure 7 shows
that the overlap reduction function for the Virgo-MiniGRAI cross correlation
is very small, but there is a maximum at about $2710Hz$. Approximately
this maximum is reached when

\begin{equation}
(\frac{f}{2710Hz})(\frac{d}{1090km})\simeq1.\label{eq: Max}\end{equation}
\begin{figure}
\includegraphics{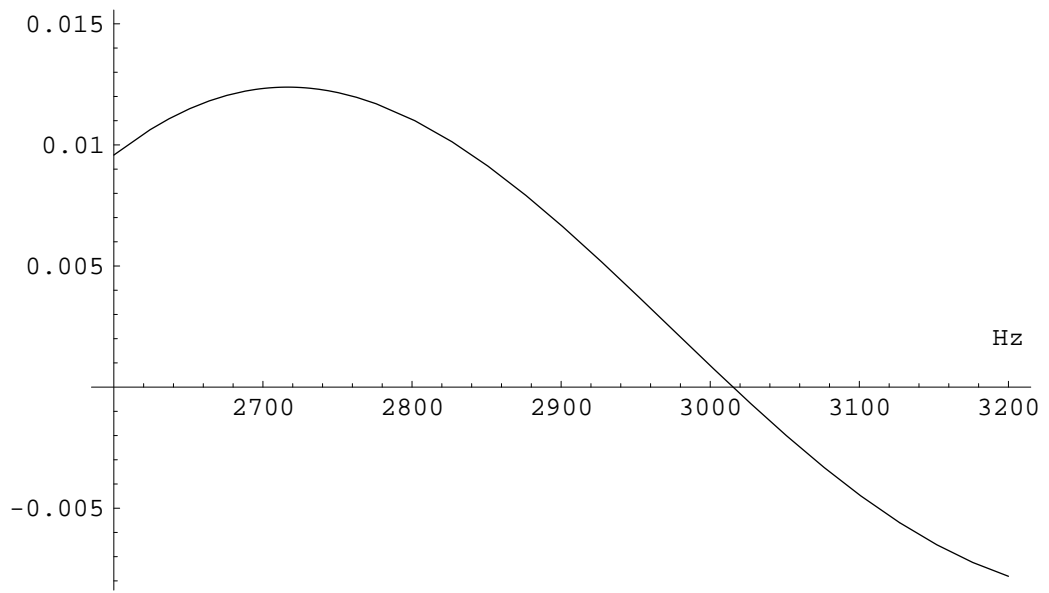}

\caption{the frequency dependence of the Virgo-MiniGRAIL overlap reduction
function in the MiniGRAIL frequency range}
\end{figure}

Reasuming, in this letter the frequency - dependent angular pattern
of interferometers in the TT gauge has been reanalyzed following \cite{key-6}.
After this, the result has been used to study a cross correlation
between the Virgo interferometer and the MiniGRAIL resonant sphere
generalizing the analysis of \cite{key-13} for a real detector. It
has been shown that the overlap reduction function for the cross correlation
bewteen Virgo and the monopole mode of MiniGRAIL is very small, but
a maximum in the correlation has also be found at about $2710Hz$,
in the range of the MiniGRAIL sensitivity.

\section*{Acknowledgements}

I would like to thank Salvatore Capozziello, Giancarlo Cella and Francesco
Rubanu for helpful advices during my work. I have to thank the European
Gravitational Observatory (EGO) consortium for the using of computing
facilities.


\begin{thebibliography}{10}
\bibitem{key-1}Acernese F et al. (the Virgo Collaboration) - Class. Quant. Grav.
\textbf{23} 19 S635-S642 (2006); 
\bibitem{key-2}Corda C - Astropart. Phys. \textbf{27,} No 6, 539-549 (2007); 
\bibitem{key-3}Hild S (for the LIGO Scientific Collaboration) - Class. Quant. Grav.
\textbf{23} 19 S643-S651 (2006) 
\bibitem{key-4}Willke B et al. - Class. Quant. Grav. \textbf{23} 8S207-S214 (2006); 
\bibitem{key-5}Tatsumi D, Tsunesada Y and the TAMA Collaboration - Class. Quant.
Grav. \textbf{21} 5 S451-S456 (2004)  
\bibitem{key-6}Capozziello S and Corda C - Int. J. Mod. Phys. D \textbf{15} 1119
-1150 (2006) 
\bibitem{key-7}Capozziello S - \textit{Newtonian Limit of Extended Theories of Gravity}
in \textit{Quantum Gravity Research Trends} Ed. A. Reimer, pp. 227-276
Nova Science Publishers Inc., NY (2005) \foreignlanguage{italian}{arXiv:}gr-qc/0412088
\bibitem{key-8}Capozziello S and Troisi A Phys. Rev. D \textbf{72} 044022 (2005) 
\bibitem{key-9}Allemandi G, Capone M, Capozziello S and Francaviglia M - Gen. Rev.
Grav. \textbf{38} 1 (2006) 
\bibitem{key-10}Allemandi G, Francaviglia M, Ruggiero ML and Tartaglia A - Gen. Rev.
Grav. \textbf{37} 11 (2005) 
\bibitem{key-11}Brunetti M, Coccia E, Favone V and Fucito F - Phys. Rew. D \foreignlanguage{italian}{\textbf{59}
044027 (1999) }
\bibitem{key-12}Tobar ME, Suzuki T and Kuroda K \foreignlanguage{italian}{Phys.} Rev.
\foreignlanguage{italian}{D 59 \textbf{}102002 (1999);}
\bibitem{key-13}Maggiore M and Nicolis A - \foreignlanguage{italian}{Phys.} Rev. \foreignlanguage{italian}{D
\textbf{62} 024004 (2000); also in gr-qc/9907055}
\bibitem{key-14}Bonasia N and Gasperini G Phys. Rew. D \textbf{71} 104020 (2005)
\selectlanguage{italian}
\bibitem{key-15}Misner CW, Thorne KS and Wheeler JA - {}``Gravitation'' - W.H.Feeman
and Company - 1973 \foreignlanguage{english}{}
\selectlanguage{english}
\bibitem{key-16}Maggiore M- Physics Reports \textbf{331}, 283-367 (2000) 
\bibitem{key-17}Rakhmanov M - Phys. Rev. D \textbf{71} 084003 (2005) 
\bibitem{key-18}De Waard A et al - Proceedings of the 6th Edoardo Amalfi Conference
ooon Gravitatinal Waves, Bankolu Shinryoukan Kise Nago, Okinawa Japan,
June 20-24, 2005
\bibitem{key-19}De Waard A et al - Classical and Quantum Gravity \foreignlanguage{italian}{\textbf{22}}
S215-S219 
\bibitem{key-20}Landau L and Lifsits E - {}``Teoria dei campi'' - Editori riuniti
edition III (1999) 
\bibitem{key-21}Flanagan E - Phys. Rev. D \foreignlanguage{italian}{\textbf{48} 2389
(1993) }
\bibitem{key-22}Babusci D, Baiotti L, Fucito F and Nagar A - Phys. Rew. D \textbf{64}
062001 (2001)
\end{thebibliography}
\end{document}